\documentclass{PoS}

\title{NNNLO results on top-quark pair production near 
threshold\thanks{
preprint PITHA~08/03, TTP/08-04, SFB/CPP-08-07}}

\ShortTitle{NNNLO results on top-quark pair production near 
threshold}

\author{\speaker{Martin Beneke}\\
        Institut f{\"u}r Theoretische Physik~E,
        RWTH Aachen, D--52056 Aachen, Germany\\
        E-mail: \email{mbenekeATphysik.rwth-aachen.de}}

\author{Yuichiro Kiyo\\
       Institut f{\"u}r Theoretische Teilchenphysik,
       Universit{\"a}t Karlsruhe, D--76128 Karlsruhe, Germany}

\author{Kurt Schuller\\
        Institut f{\"u}r Theoretische Physik~E,
        RWTH Aachen, D--52056 Aachen, Germany}

\abstract{We present new results on the NNNLO top-antitop 
production cross section near threshold from potential and 
ultrasoft gluon corrections. The new non-logarithmic 
third-order terms are 
in the 10\% range and lead to a significant reduction in 
the theoretical error.}

\FullConference{8th International Symposium on Radiative Corrections \\
                 October 1-5, 2007\\
                 Florence, Italy}

\begin{document}

\section{Introduction}

High-order perturbative calculations of non-relativistic heavy 
quark-antiquark systems are required for precise quark mass 
determinations from QCD sum rules or the lowest upsilon states 
(bottom quark mass) or the energy dependence of the threshold top-quark pair
production cross section in $e^+ e^-$ collisions (top quark mass). 
Concerning the $t\bar t$ quark cross section the status is 
as follows: the NNLO calculations performed about ten years ago 
revealed a large uncertainty, up to $\pm 25\%$, in the cross 
section in the resonance peak 
region \cite{Hoang:2000yr,Beneke:1999qg}. Subsequent 
calculations that include a summation of logarithms of 
$\alpha_s$ find a much reduced scale dependence 
around $\pm (3-6)\%$ \cite{Hoang:2001mm,Hoang:2003xg}. The 
main effect comes from the logarithms at the third order (NNNLO) 
rather than the all-order series \cite{Pineda:2006ri}. 
Since at NNNLO the ultrasoft scale $m_t\alpha_s^2\sim 2\,
\mbox{GeV}$ appears for the first time, a complete calculation 
of the (non-logarithmic) NNNLO correction is needed. Similar 
conclusions apply to bottomonium systems, with larger 
uncertainties.

Recently the NNNLO correction to the $S$-quarkonium wave-functions 
at the origin (corresponding to the residues of the poles of the 
heavy quark current spectral functions) from potential insertions 
and ultrasoft gluons have been 
completed~\cite{Beneke:2007gj,Beneke:2007pj}. The talk presented 
at this conference summarized these results together with 
NNNLO results on the full energy-dependent spectral function \cite{inprep}
relevant to the $t\bar t$ cross section. Since the combined 
result of~\cite{Beneke:2007gj,Beneke:2007pj} has already been 
discussed in another proceedings article \cite{Beneke:2007uf}, 
we focus here on the full spectral function and the case of 
the top quark. That is, we consider the two-point function
\begin{eqnarray}
\label{eq:TwoPointFunction}
&&
\left(q^\mu q^\nu - g^{\mu\nu} q^2 \right)\Pi(q^2)
= i \int d^d x\, e^{i q x}\,
\langle \Omega | T(j^\mu(x)j^\nu(0)) |\Omega\rangle
\end{eqnarray}
of the electromagnetic top-quark current $j^\mu=\bar{t}\gamma^\mu
t$, choosing $q^\mu=(2 m_t+E,\vec{0})$ with $m_t$ the
pole mass of the top quark and $E$ of order of a few GeV. The 
width of the top quark is taken into account by simply letting 
$E\to E+i\Gamma_t$ become complex \cite{FK87}. However, one should 
note that a fully consistent treatment of the top-quark decay 
beyond the NLO approximation 
requires the inclusion of many other electroweak effects that 
are not yet known.

\section{Remarks on the calculation}

After integrating out the hard and soft momentum scales the problem 
is reduced to the calculation of a non-relativistic correlation 
function $G(E)$ to third order in non-relativistic perturbation 
theory. The perturbations consist of potential insertions
(instantaneous, spatially non-local operators) and ultrasoft 
gluon interactions with the top quarks. Since an infinite number of 
potential (Coulomb) gluons can be exchanged between the heavy quarks 
without parametric suppression, 
the free heavy quark-anti-quark propagators are promoted to the 
Green function of the Schr\"odinger operator $H_0=-\vec{\nabla}^2/m_t
-(\alpha_s C_F)/r$ with the colour 
Coulomb potential ($C_F=4/3$). Computing 
Feynman integrals with Coulomb Green functions while simultaneously 
regulating all divergences dimensionally to be consistent with 
fixed-order matching calculations is the main challenge of the 
NNNLO calculation. The third-order correction to $G(E)$ is composed of
\begin{eqnarray}
\delta_3 G &=&  - \langle 0|\hat{G}_0 \delta V_1 \hat{G}_0
\delta V_1\hat{G}_0 \delta V_1 \hat{G}_0|0\rangle + 
2\, \langle 0|\hat{G}_0 \delta V_1 \hat{G}_0 \delta V_2
\hat{G}_0|0\rangle - \langle 0|\hat{G}_0\delta V_3 \hat{G}_0
|0\rangle  + \delta G_{\rm us},\qquad
\label{git}
\end{eqnarray}
where $\hat{G}_0= (H_0 - E-i \epsilon)^{-1}$, 
$|0\rangle$ denotes a relative position eigenstate with 
eigenvalue ${\bf r}=0$, and  $\delta V_i$ the $i$th order 
perturbation potentials (see \cite{Beneke:2007gj}).  The 
contributions to $G(E)$  involving only higher-order corrections to 
the Coulomb potential have already been computed in \cite{Beneke:2005hg}
and are included in the following numerical result. 
The ultrasoft contribution is ($D = d-1$)
\begin{eqnarray}
\delta G_{\rm us} &=&
(-i) (i g_s)^2 C_F \int\frac{d^d k}{(2\pi)^d}\,
\frac{-i}{k^2}\left(\frac{k^i k^j}{k_0^2}-\delta^{ij}\right)
\int \prod_{n=1}^6 \frac{d^{D}p_n}{(2\pi)^{D}}\,\;
i \tilde G^{(1)}(p_1,p_2;E)
\nonumber\\
&& \hspace*{0cm}
\times \,i\left[\frac{2 p_3^i}{m_t}
(2\pi)^{D}\delta^{(D)}(p_3-p_2)+(i g_s)^2 \frac{C_A}{2} 
\frac{2 (p_2-p_3)^i}{(p_2-p_3)^4}\right]
\,i \tilde G^{(8)}(p_3,p_4;E+k^0)
\nonumber\\
&& \hspace*{0cm}
\times \,i\left[-\frac{2 p_4^j}{m_t}
(2\pi)^{D}\delta^{(D)}(p_4-p_5)+(i g_s)^2 \frac{C_A}{2} 
\frac{2 (p_4-p_5)^j}{(p_4-p_5)^4}\right]
\,i \tilde G^{(1)}(p_5,p_6;E)
\end{eqnarray}
with $\tilde G^{(1,8)}(p,p^\prime;E)$ the colour-singlet/octet 
momentum-space Coulomb Green functions. In position space 
this expression simplifies to three instead of seven loop 
integrations and similar simplifications apply to the other 
terms in (\ref{git}). However, the integrals are divergent 
and the $1/\epsilon$ poles must be extracted in momentum 
space. A further complication is that the Coulomb Green 
functions are not known in $D$ dimensions. The strategy 
therefore consists of identifying all divergent 
subgraphs, and to calculate them in $d$-dimensional momentum 
space. Then combine the result with the sub-divergence 
counterterms related to the renormalization of potentials 
and non-relativistic currents and perform the remaining 
integrations in three dimensions. Due to the incomplete 
treatment of finite width effects an over-all divergence 
$\alpha_s^{1,2}/\epsilon\times \Gamma_t$ remains in the 
$t\bar t$ cross section, which is minimally subtracted in the 
result below.

\section{Size of logarithmic and non-logarithmic 
terms}

We consider the residue of the correlation function 
at the lowest-energy bound-state pole at 
$E_1=-m_t (\alpha_s C_F)^2/4+\ldots$ to compare the new NNNLO 
non-logarithmic terms~\cite{Beneke:2007gj,Beneke:2007pj} to 
the previously known logarithms \cite{KniPen2,ManSte,KPSS2,Hoa2}, 
since in this case a simple numerical result can be given. 
$Z_1$, defined by 
\begin{equation}
\Pi(q^2) \stackrel{E\rightarrow E_1}{=}
\frac{3}{2 m_t^2}\,\frac{Z_1}{E_1-E-i \epsilon},
\end{equation}
is related to the height of the cross section peak by
$R_{\rm peak} \approx 18\pi e_t^2 Z_1/(m_t^2\Gamma_t)$, 
so we expect similar conclusions to hold for the entire 
$t\bar t$ cross section. The NNNLO expression for $Z_1$ 
reads
\begin{eqnarray}
Z_1 &=& \frac{(m_t \alpha_s C_F)^3}{8\pi}
\times \bigg( 1+
\alpha_s \Big[-2.13+3.66\,L\Big] 
\nonumber \\
&& +\, 
\alpha_s^2 \Big[8.38-7.26 \,\ln\alpha_s -13.40 \, L 
+ 8.93 \,L^2\Big]
\nonumber\\
&& +\, \alpha_s^3\Big[11.01 + [37.58]_{c_3,n_f} -9.79  \,\ln\alpha_s 
-16.35  \,\ln^2\alpha_s 
\nonumber\\
&& \hspace*{0.7cm} +\,(53.17- 44.27 \,\ln\alpha_s) \,L 
- 48.18 \,L^2+ 18.17 \,L^3\Big]
\bigg)
\nonumber\\
&=& \frac{(m_t \alpha_s C_F)^3}{8\pi}
\times \bigg( 1-2.13 \alpha_s +22.64 \alpha_s^2 + 
[-32.96+[37.58]_{c_3,n_f}]\alpha_s^3\bigg). 
\label{Z1}
\end{eqnarray}
Since the scales $m_t$, $m_t\alpha_s$ and $m_t\alpha_s^2$ are all
relevant here, there is an ambiguity in the representation of the 
logarithms. The above result uses $\alpha_s\equiv \alpha_s(\mu)$ 
and puts the explicit $\mu$-dependence into 
$L\equiv\ln \mu/(m_t C_F\alpha_s)$. The NNNLO result is 
not yet complete: the constant 11.01 includes an estimate 
of the NNNLO correction to the Coulomb potential, 
$a_3=3840$ \cite{Chishtie:2001mf}, and sets certain unknown 
${\cal O}(\epsilon)$ potential terms to zero. This is expected 
to have a minor effect \cite{Beneke:2007gj}. More important 
is that only the $n_f$-parts of the third-order matching 
coefficient $c_3$ of the non-relativistic current $\psi^\dagger
\sigma^i\chi$ are known \cite{Marquard:2006qi}, which turn out 
to be very large ($[37.58]_{c_3,n_f}$). In the following numerical 
results for the $t\bar t$ cross section we therefore consider 
two options, one where the constant part of $c_3$ is set 
to the known $n_f$ terms, the other where it is set to zero. 
(The logarithms are all known and always included.)

We observe (third and fourth line of (\ref{Z1})) that the  
typical size of non-logarithmic terms of individual third-order 
corrections (ultrasoft, non-Coulomb potentials, Wilson coefficient) 
is about $40\alpha_s^3 \approx 10\% \,\,(\,>100\%)$ 
for toponium (bottomonium). However, large cancellations 
between individual terms and between logarithmic 
and non-logarithmic terms occur. Thus, to obtain a reliable 
third-order result the non-logarithmic terms are crucial 
and a final assessment needs the missing $n_f$-independent 
term in $c_3$. This is seen in the last line 
of (\ref{Z1}), which shows the series for $\alpha_s=0.14$ 
where $L=0$. The large NNLO correction is evident. On the 
other hand, the third-order correction is not anomalously 
large, although the final coefficient will only be known 
when the term $[37.58]_{c_3,n_f}$ is replaced by the full result 
for $c_3$. The NNNLO result shows a strong reduction of 
the scale dependence compared to NNLO as discussed  
in \cite{Beneke:2007uf}, but the perturbative prediction becomes 
unstable for $\mu<20\,\mbox{GeV}$. A study of this problem 
for the Coulomb corrections, where a resummation of the perturbative 
series can be done by means of a numerical solution, has shown 
\cite{Beneke:2005hg} that the perturbative prediction for  
$\mu>25\,\mbox{GeV}$ is close to the true result, hence 
we do not consider scales $\mu<25\,$GeV.

\section{Top-quark cross section}

We next discuss the $t\bar t$ production cross section in 
$e^+ e^-$ annihilation near threshold. More precisely, we 
consider the $R$-ratio
\begin{equation}
R = \sigma_{t\bar tX}/\sigma_0 = 12\pi e_t^2 \,\mbox{Im}\;\Pi(q^2) 
\quad\qquad \left(\sigma_0=4\pi\alpha_{\rm
  em}^2/(3 s)\right),
\label{R}
\end{equation}
neglecting the axial-vector contribution from $Z$-exchange for 
the purpose of discussing the impact of the QCD NNNLO correction. 
The top quark pole mass should be avoided as an input parameter. 
Here we use the potential-subtracted mass \cite{Beneke:1998rk} 
implemented as explained in \cite{Beneke:2005hg}. The parameters for 
the cross section calculation are: 
$m_{t,\rm PS}(20\,\mbox{GeV})=175\,\mbox{GeV}$, 
$\Gamma_t=1.4\,$GeV, $\alpha_s(M_Z)=0.1189$.

\begin{figure}[t]
  \begin{center}
  \includegraphics[width=0.57\textwidth]{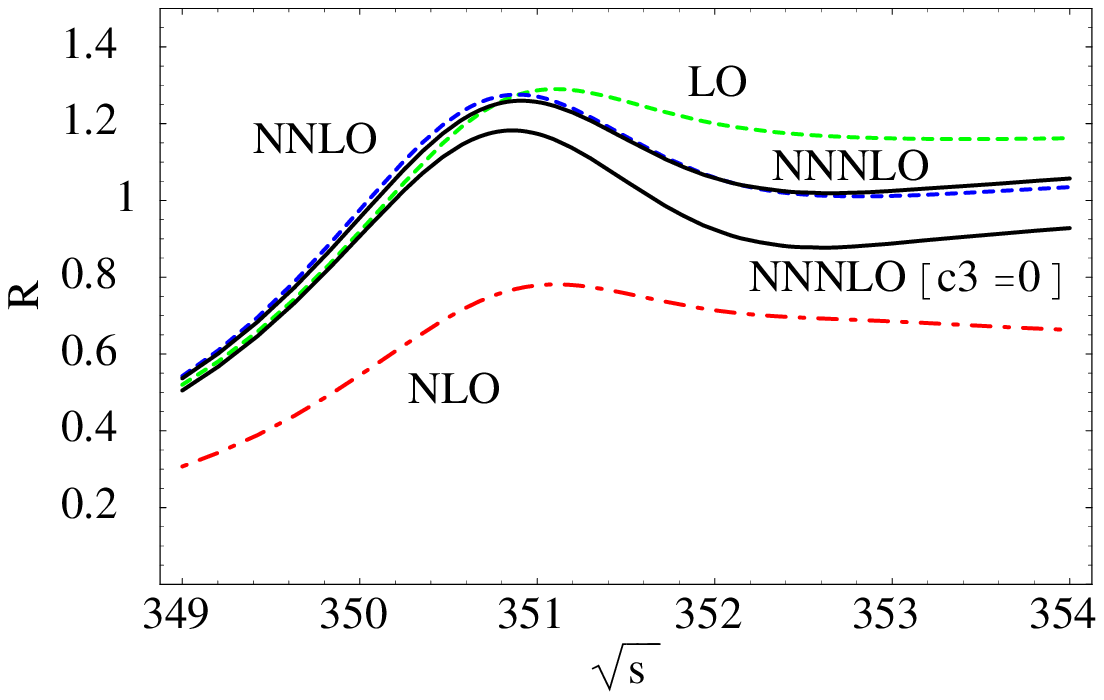}
  \vskip0.5cm
   \includegraphics[width=0.58\textwidth]{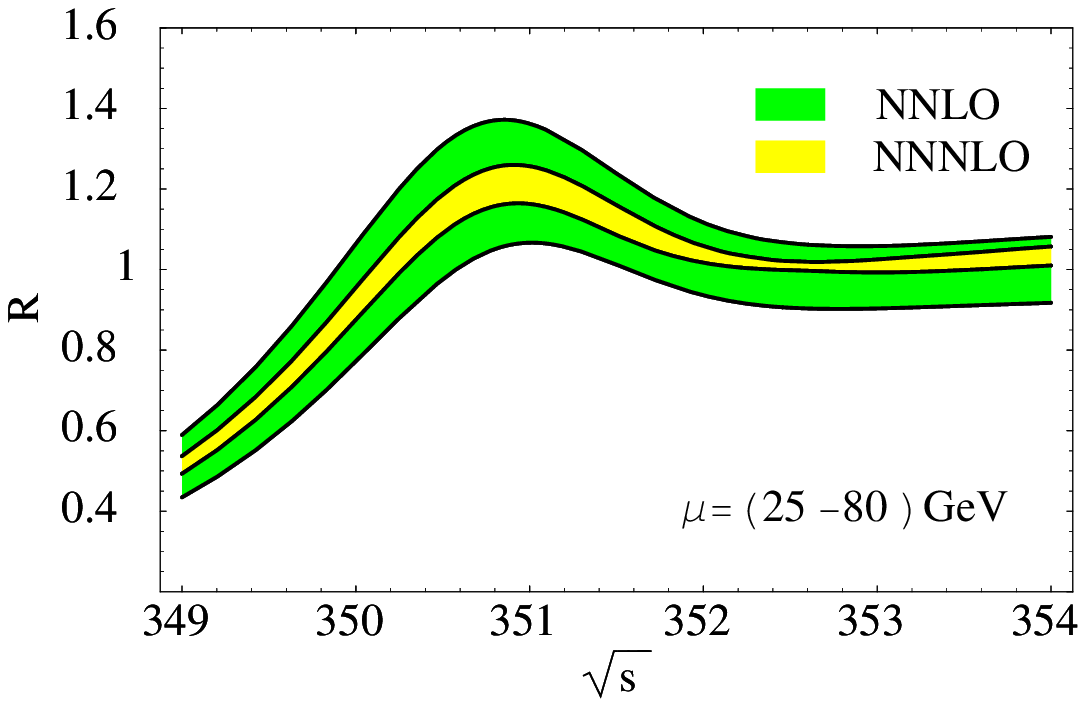}
  \caption{(top) Successive approximations to the $R$-ratio 
  at fixed $\mu=30\,\mbox{GeV}$. At NNNLO two implementations of 
  $c_3$ are shown as discussed in the text. (bottom) 
  Renormalization scale dependence at NNLO and NNNLO. Here 
  all known terms in $c_3$ are included.}
  \label{fig1}
  \end{center}
\end{figure}

The successive LO ... NNNLO approximations to $R$ are shown 
in Figure~\ref{fig1} (top panel). At $\mu=30\,\mbox{GeV}$ 
the size of the third-order correction is up to 10\% 
depending on the assumption for $c_3$. When all known 
terms in $c_3$ are included the peak cross section is 
about 10\% larger than in the renormalization-group improved 
NNLO calculations \cite{Hoang:2001mm,Hoang:2003xg,Pineda:2006ri} 
due to the sizeable constant term related to 
$11.01 + [37.58]_{c_3,n_f}$ in (\ref{Z1}). Contrary to 
the NNLO approximation the third-order result shows good 
convergence of the perturbative expansion. The bottom panel of 
Figure~\ref{fig1} consequently displays a strong reduction 
of the scale dependence from NNLO to NNNLO. 
The residual scale dependence is 
similar at NNNLO and in the renormalization-group improved 
calculations, which already captures correctly 
the logarithms of $\mu$. It therefore appears that with a 
complete NNNLO result and a summation of higher-order 
logarithms at hand, the demands on an accurate theoretical 
prediction of the cross section near threshold can be met 
{\em as far as QCD corrections are concerned.} In particular, 
the scale dependence of the peak position which is indicative 
of the accuracy of the top mass measurement is now well 
below 100 MeV.

\section{Summary}

The NNNLO QCD correction to the $t\bar t$ cross section 
near threshold is now nearly complete. We presented for the 
first time the result of the third-order potential and 
ultrasoft correction to the non-relativistic heavy-quark correlation 
function. We find that the third-order correction 
behaves well (contrary to the anomalously large effect at NNLO) 
and removes a large part of the theoretical 
uncertainty. The new non-logarithmic terms are numerically 
important and increase the cross section relative to the 
renormalization-group improved NNLO result by about 10\%, 
when all presently 
known terms of the three-loop matching coefficient $c_3$ 
are included. Further work is necessary on a consistent 
treatment of electroweak and finite-width effects  
(see \cite{Beneke:2003xh,Hoang:2004tg,Beneke:2007zg}).

\section*{Acknowledgments}
This work was supported by the DFG Sonderforschungsbereich/Transregio
9 ``Computer-gest\"utzte Theoretische Teilchenphysik'' and DFG
Graduiertenkolleg ``Elementarteilchen\-physik an der TeV-Skala''.

\end{document}